\documentstyle[preprint,eqsecnum,prd,aps]{revtex}
\begin{document} 

\input{epsf}
\def\Journal#1#2#3#4{{#1} {\bf #2}, #3 (#4)}

\def\NCA{\em Nuovo Cimento}
\def\NIM{\em Nucl. Instrum. Methods}
\def\NIMA{{\em Nucl. Instrum. Methods} A}
\def\NPB{{\em Nucl. Phys.} B}
\def\PLB{{\em Phys. Lett.}  B}
\def\PRL{\em Phys. Rev. Lett.}
\def\PRD{{\em Phys. Rev.} D}
\def\ZPC{{\em Z. Phys.} C}

\preprint{
\vbox{
\halign{&##\hfil\cr
	& ANL-HEP-PR-97-42 \cr
	& July 10, 1997 \cr}}
}
\title{Prompt Photon Plus Jet Photoproduction at HERA at Next-to-Leading
Order in QCD}

\author{L. E. Gordon}
\address{High Energy Physics Division, Argonne National Laboratory,
	Argonne, IL 60439}
\maketitle
\begin{abstract} 
The cross section for photoproduction of an isolated prompt photon in 
association with a jet is studied in Next-to-Leading Order. The
kinematics are those appropriate for the DESY $ep$ 
collider HERA. The effects on the cross section of various experimental cuts  
including isolation cuts on the photon is examined. Comparisons with the ZEUS 
preliminary data using two parametrizations of the photon structure 
function is made, and good agreement is found. The data is not yet 
precise enough to make a distinction between various models for the photon 
structure function.  
\vspace{0.2in}
\end{abstract}
\narrowtext

\section{Introduction}

It has long been anticipated that the DESY $ep$ collider HERA would
provide a good opportunity to study prompt photon production in
photoproduction processes \cite{aur1}. Over the past few years various
calculations of this process have been performed leading to continuous 
improvements in their theoretical precision \cite{bks,afg,gs,gv1}. In the most
recent studies \cite{gv1,gv2} the inclusive cross section for producing
a single photon was calculated fully in NLO with photon isolation effects
incorporated. Gordon and Vogelsang \cite{gv1,gv2} use an approximate but 
nevertheless accurate analytic technique \cite{gv3,ggrv} for including 
isolation effects in the NLO calculation, including the fragmentation 
contributions. This analytic technique is only applicable to single inclusive 
prompt photon production and cannot be applied when a jet is also observed.

The ZEUS Collaboration have reported prompt photon data
\cite{zeus} and have first chosen to analyse events with a jet
balancing the transverse momentum ($p_T^{\gamma}$) of the photon. In order to
compare with this data a new calculation is necessary as
described in outline in the next section.

In all previous studies of prompt photon production at HERA, one of
the common themes was the possibility of using it for measuring the 
photon distribution functions, particularly the gluon distribution,
$g^{\gamma}(x,Q^2)$ which is presently poorly constrained by the available
data. This latter fact is still true even with the availability of jet 
photoproduction data from both HERA and TRISTAN. Prompt 
photon production is particularly attractive since it is dominated in Leading
Order (LO) by the hard scattering subprocess $q g\rightarrow \gamma q$,
resulting in a cross section which is very sensitive to the gluon
distribution. 

At HERA the situation is more complicated than at hadron
colliders for two reasons. Firstly there are two particles involved in the
reaction, namely the quasi-real photon (emitted by the electron which 
scatters at a small angle) and the proton. Both particles have
distinct gluon distribution functions $g^{\gamma}$ and $g^p$, hence two
different $qg$ initiated subprocess are present,
$q^p g^{\gamma}\rightarrow \gamma q$ and  $q^{\gamma} g^p\rightarrow
\gamma q$. Since they contribute to the cross section in different
regions of pseudo-rapidity, $\eta$, it has been proposed that this may provide 
a means of separating them, but this has proven to be 
difficult to implement in the experiments. Secondly, there are two 
contributions
to the cross section in photoproduction processes, usually labelled the direct
and resolved. In the former case the quasi-real photon
participates directly in the hard scattering subprocess and gives up all
its energy, while in the latter, resolved, case it interacts via its
partonic substructure. Thus the resolved subprocesses are sensitive to
the photon structure functions whereas the direct are not. Again it was
proposed that they may be separated experimentally with suitable
rapidity cuts, but these studies  assumed a fixed initial
photon energy. Since the
initial photon energy is not fixed but forms a continuous spectrum, then
even this separation is not straightforward \cite{gs}. This is because the 
spectrum of
initial photon energies causes the sharply separated peaks in the
rapidity spectrum of the resolved and
direct components, present when the initial photon energy is fixed,
to become smeared out and so less sharply defined. Separation of the resolved 
and direct processes is better achieved by tagging of the spectator jet from
the resolved photon.   
  
In section II a brief outline of the theoretical background to the cross
section as well as the technique of calculation is given. In section III
numerical results are presented and in section IV the summary and
conclusions are presented. 

\section{The Inclusive Photon Plus Jet Cross Section}

\subsection{Contributing Subprocesses}

In addition to the direct and resolved photon contributions to the cross
section there are the non-fragmentation and fragmentation contributions.
In the former case the observed final state photon is produced directly in the 
hard scattering whereas in the latter it is produced by long distance
fragmentation off a final state parton. The fragmentation processes
involve the functions which cannot be calculated and must be taken
from experiment. So far they have not been satisfactorily
measured. There are various parametrizations of these
functions available using different models for the input
distributions. As the numerical results will show in
the next section these contributions are small at HERA energies and so do
not provide a significant source of uncertainty in the present calculation.
This point has already been noted in previous studies and will be
returned to below.

The only direct non-fragmentation process contributing to the cross section in 
LO is the so called QCD Compton process (fig.1a)
\begin{displaymath} 
q \gamma \rightarrow \gamma q.
\end{displaymath}
The corresponding direct fragmentation processes in LO (fig.1b)are 
\begin{displaymath}
q \gamma \rightarrow g q\;\;\; {\rm and}\;\;\; g \gamma \rightarrow q\bar{q}.
\end{displaymath}
As discussed in many places (see eg., \cite{gv1}) the photon fragmentation
function is formally $O(\alpha_{em}/\alpha_s)$, thus although the hard
subprocess cross sections in the fragmentation case are
$O(\alpha_{em}\alpha_s)$, after convolution with the photon fragmentation
functions the process contributes at $O(\alpha_{em}^2)$, the same as the the
non-fragmentation part. Thus in a fixed order calculation the two
contributions must be
added together to provide the physical cross section. 

At NLO for the non-fragmentation part there are the virtual corrections
to the LO Compton process plus the additional three-body processes
\begin{displaymath}
q \gamma \rightarrow \gamma g q\;\;\; {\rm and}\;\;\; g \gamma \rightarrow
\gamma q\bar{q}.
\end{displaymath}
These processes have been calculated previously by various authors. In
this study the virtual corrections are taken from \cite{gvx}  
In addition there are $O(\alpha_s)$ corrections to the fragmentation
processes to take into account, but in this calculation these processes
are included in LO only. It has been shown previously \cite{gv1} that the
fragmentation contributions are not as significant here as at hadron
colliders which generally have higher cms energies. They are also reduced 
drastically 
when isolation cuts are implemented. Thus ignoring NLO corrections to the
fragmentation contributions, while in principle theoretically
inconsistent, will 
not lead to significant error in estimates of the cross section.

In the resolved case, for non-fragmentation there are only the two
processes  
\begin{displaymath}
 q g\rightarrow \gamma q\;\;\; {\rm and}\;\;\; q \bar{q}\rightarrow \gamma g.
\end{displaymath}
in LO (fig.2). At NLO there are virtual and three-body corrections to these 
as well as 
other three-body processes, for example, $g g\rightarrow \gamma q \bar{q}$ 
etc. For a complete list of these plus the fragmentation processes see,
for example, ref.\cite{gv1}.
As with the direct case, the fragmentation contributions are included
here in LO.

\subsection{Some Calculational Details}

The calculation was performed using the phase space slicing method which
makes it possible to perform photon isolation exactly as well as to
implement the jet definition in the NLO calculation. More details of parts of 
the calculation can be 
found in ref.\cite{gordon2}. The two-body matrix elements for the
resolved case, after the soft and 
collinear poles have been canceled and factorized in the $\overline{MS}$ 
scheme can be found in the appendices of refs.\cite{gordon2,boo}. Those for the direct contributions can be obtained from these 
by appropriately removing non-abelian couplings. These matrix elements depend 
on the soft and collinear cut-off parameters, $\delta_s$ and $\delta_c$ and 
must be added to the three-body matrix elements, also included in the appendix 
of \cite{gordon2}, in order to cancel the dependence of the cross section
on these arbitrary cut-off parameters. 

Following the ZEUS experiment, the cone isolation method is used to
isolate the photon signal. This method restricts the hadronic energy allowed 
in a cone of radius $R_{\gamma}=\sqrt{\Delta \phi^2 + \Delta \eta^2}$,
centred on the photon to be
below the value $\epsilon E_{\gamma}$, where $E_{\gamma}$ is the photon
energy. The fixed value $\epsilon= 0.1$ is used in this study, which 
corresponds to the value used in the ZEUS analysis. By contrast the CDF
collaboration in their analysis \cite{cdf} uses a value of $\epsilon = 2\;{\rm
GeV}/p_T^{\gamma}$, which varies with the photon energy ($p_T^{\gamma}$
is the transverse momentum of the observed photon).  

The cone algorithm is also used to define the jet. This defines a jet as
hadronic energy deposited in a cone radius $R_J=
\sqrt{\Delta \phi^2 + \Delta \eta^2}$. If two partons form the jet then
the kinematic variables are combined to form that of the jet according to the
formulae 
\begin{eqnarray}
p_J&=&p_1+p_2 \nonumber \\
\eta_J&=& \frac{(\eta_1 p_1 + \eta_2 p_2)}{p_1+p_2} \nonumber \\
\phi_J&=& \frac{(\phi_1 p_1 + \phi_2 p_2)}{p_1+p_2}.
\end{eqnarray} 
In the ZEUS analysis $R_{\gamma}=1.0$ and $R_J=1.0$ are chosen and 
these values will also be used in this study.

In order to estimate the flux of quasi-real photons from the electron
beam the Weiszacker-Williams approximation is used. Thus the `electron
structure function' $f_e(x_e,Q^2)$ is given by a convolution of the
photon structure function $f^{\gamma}(x_{\gamma},Q^2)$ and the
Weiszacker-Williams (WW) function
\begin{eqnarray}
f_{\gamma/e}(z)&=&\frac{\alpha_{em}}{2\pi}\left[\left\{
\frac{1+(1-z)^2}{z}\right\} \ln\frac{Q^2_{max}(1-z)}{m_e^2 z^2} 
\right. \nonumber \\
& - & \left. 2 m_e^2
z \left\{ \frac{(1-z)}{m_e^2 z^2}-\frac{1}{Q^2_{max}}\right\} \right]
\end{eqnarray}
by
\begin{equation}
f_e(x_e,Q^2)=\int^1_{x_e}\frac{dz}{z}f_{\gamma/e}(z)f^{\gamma}\left(\frac{x_e}
{z},Q^2\right).
\end{equation}
The expression for $f_{\gamma/e}(z)$ was taken from ref.\cite{gas}.
Following the ZEUS analysis the value $Q^2_{max}=1$ GeV$^2$ is used
throughout. 

\section{Results}

\subsection{Effect of Experimental Selections}

The numerical results presented in this section are obtained using the
GS96 \cite{gs96} photon distribution functions, the CTEQ4M \cite{cteq} parton
distributions for the proton and the GRVLO \cite{grvf} fragmentation functions 
as standard. Futhermore the two-loop expression for $\alpha_s$ is used, 
four-flavours of quarks are assumed active and the
factorization/renormalization scales are taken to be equal to the photon
$p_T$ ($Q^2=(p_T^{\gamma})^2$). The maximum virtuality of the initial
state photon is fixed at $Q^2_{max}=1$ GeV$^2$. The calculation is performed 
in the $ep$ laboratory frame
using $P_e=27.5$ GeV for the electron energy and $P_p=820$ GeV for the
proton energy. The electron is moving toward negative rapidity.

In order to make contact with the results of previous calculations, it
is convenient to start by examining the inclusive single prompt photon cross
section, $ep\rightarrow \gamma X$. As more data are taken at HERA this
cross section (with isolation cuts) will certainly be measured since it
is the largest cross section involving prompt photon production.
In fig.3a the non-isolated single inclusive
prompt photon cross section is shown as a function of photon rapidity 
at $p_T^{\gamma}=5$ GeV. No experimental cuts are implemented. 
In the positive
rapidity region the resolved contributions are roughly twice as large as the
direct and thus this is the region of interest if information on the
gluon distribution of the photon is to be obtained. At negative rapidity,
the direct and resolved contributions are comparable in size. 

When the WW spectrum is cut as done by the ZEUS
Collaboration ($0.16\leq z \leq 0.8$) the cross section changes as shown in
fig.3b (also at the same $p_T^\gamma= 5$ GeV). Both the resolved and direct 
contributions remain essentially unchanged at negative rapidities but 
are reduced in the positive rapidity region. The effect on the direct 
contribution is large, being reduced by a factor of $10$ at
$\eta^{\gamma}=2$. Thus sensitivity to the photon structure function
is enhanced in this region since the resolved contribution does not fall
by as much. The reason for the asymmetric
response of the two contributions to this cut is that the WW
distribution is largest at small-$z$ ($x_e=z$, for the direct events). 
Cutting out this region
removes a large fraction of the direct events with lower energy initial
photons. When the convolution in eq.(2.3) is
taken for the resolved processes on the other hand, for a given
$x_{\gamma}=x_e/z$, all regions of $x_e$ contribute and thus the
cut on $z$ does not have the same dramatic effect in this region. In all
the following results the cut on $z$ is implemented.

Using the standard parameters, the fragmentation contribution 
constitutes less than $20\%$ of the cross section at $p_T^{\gamma}=5$ GeV 
(before isolation) and as expected, falls rapidly with increasing 
$p_T^{\gamma}$. After
isolation, the fragmentation contribution is reduced to about $3\%$ of the
cross section. Fig.3c shows the
contribution from fragmentation processes to the resolved and direct
contributions, as well as their sum, at $p_T^\gamma=5$ GeV before
isolation cuts are implemented.

The higher order corrections, enhance the cross section 
by $O(20\%)$ before isolation. As indicated by fig.3d, the corrections
are numerically more significant in the positive rapidity region, but
they are still modest, indicating good perturbative stability for the
predictions. 

In Fig.4 the single inclusive prompt photon cross section at 
$p_T^\gamma=5$ GeV, with only the
cut $0.16\leq z \leq 0.8$, is compared to the photon plus jet cross
section with isolation cuts and jet definition incorporated as done by
the ZEUS collaboration. The rapidity and $p_T$
cuts $-1.5\leq \eta^J \leq 1.8$ and $p_T^J\geq 5$ GeV are placed on the
jet. As expected, the photon plus jet cross section is significantly
smaller than the single photon cross section, but does not show much
difference in shape. It could thus still potentially be used to measure
the photon distributions in the positive rapidity region.

The lower dot-dashed in fig.4 is the resolved contribution to the photon
plus jet cross section after the further cut $x_{\gamma}\geq 0.8$ is
imposed. This cut essentially removes most of the resolved contribution
to the cross section and therefore most of the sensitivity to the photon
distribution functions. It is still nevertheless not a pure direct
sample and as seen in fig.5a, it still shows sensitivity to the photonic
parton distributions. One of the main differences in the GRV and GS96
photon distributions is in the quark distributions at large-$x_\gamma$. 
In fig.5a the rapidity distribution is plotted at $p_T^\gamma=5$ GeV
with all the cuts used in the ZEUS analysis implemented, including the
cut on $x_\gamma$. At negative rapidities the photonic quark
distributions are probed at large-$x$ which is where the largest
differences between the results of GS96 and GRV are seen. By contrast,
as fig.5b demonstrates, there is almost no differences between the
results when the proton distributions are changed. This cross section
may thus potentially be used to distinguish between these two models of
the photon structure function.  

In fig.6 the cross section is plotted vs $p_T^{\gamma}$ with the ZEUS
rapidity cuts on the photon imposed ($-0.7\leq \eta_{\gamma}\leq 0.8$). 
It shows the well known fact, common to this type of photoproduction process, 
that the resolved contribution only competes with the direct at low
values of $p_T^\gamma$, while the direct dominates as $p_T^\gamma$ is
increased. One thus needs to look in the lower $p_T^\gamma$ region if
sensitivity to the photon structure function is desired and look at
higher $p_T^\gamma$ if the aim is to eliminate the resolved events. 

Fig.7 shows a partial breakdown of the isolated photon plus jet cross section
into initial state contributions as a function of $\eta^\gamma$. The
photon $p_T$ is integrated between $5$ and $10$ GeV as done by the ZEUS
Collaboration. The solid curve is the sum, the dot-dashed curve the
resolved and the dashed curve the direct. The contributions to the
resolved process are the labelled dotted curves. The dotted curve with
error bars is the $g^\gamma q^p$ initiated process as predicted using
the GRV photon distributions. Clearly it is only distinguishable from
the GS96 result in the far positive rapidity region. All other features of
the curves except for the absolute sizes of the contributions are similar 
to the results of previous studies done on single non-isolated prompt 
photon production in the $ep$ laboratory frame \cite{bks,gs,gv1}. 

\subsection{Comparison with HERA Data}

Table 1 lists predictions for the resolved and direct contributions to
the cross section and their sum for various choices of parameters. 
As stated above, in
order to obtain a sample of direct events the ZEUS Collaboration have
imposed the cut $x_{\gamma}\geq 0.8$ on their data. This cut which is
also imposed on the results in Table 1, favours the direct contributions
since they contribute at $x_{\gamma}=1$, but there is still a
contribution from the resolved processes and hence some sensitivity to
the photon distributions chosen. In addition the cuts $5$ GeV
$\leq p_T^{\gamma} \leq 10$ GeV, $p_T^J\geq 5$ GeV, $-1.5\leq \eta^J\leq
1.8$, $-0.7\leq \eta^{\gamma}\leq 0.8$ and $0.16\leq z=E_{\gamma}/E_e\leq 0.8$ 
along with the isolation
cuts and jet definitions discussed in Section II are imposed. 

The first column of numbers gives the results for the standard choice of 
parameters, while the 2nd and 3rd columns show the effect of changing the 
scales. The results
show a remarkable stability to scale changes. This is in contrast to,
for example,
the $p_T^{\gamma}$ distribution which generally shows significant scale
sensitivity. The 4th and 5th columns show the effect of changing the photon
and proton distribution functions used respectively. In the latter case,
as already indicated by the results shown in figs.5a and 5b
there is hardly any changes in the predictions, while in the former case
the changes are very significant. Since with these cuts the cross
section is mostly sensitive to the quark distributions in the photon at
large-$x$ then this measurement may potentially be used to discriminate between
the GS96 and GRV photon parametrizations which differ most significantly in
this region. The preliminary experimental value given by the ZEUS
Collaboration of $15.3\pm 3.8\pm 1.8$ pb agrees well with the NLO
theoretical predictions but the errors are still too large to make any
distinction between GS and GRV. 

\section{Conclusions}

A NLO calculation of isolated single photon plus jet production at HERA was
presented. The effects of various experimental cuts on the cross section
was studied in some detail, and comparisons are made with the preliminary data 
from the ZEUS Collaboration where good agreement was found. The kinematic cuts 
chosen favour the direct contribution but there is still a significant
sensitivity to the quarks distributions in the photon at large-$x_\gamma$. At
the moment the error in the data is still too large to distinguish
between the GRV and GS96 photon distributions, but it is expected that
analysis of more data will soon remedy this situation.  

\section*{Acknowledgments}
I am grateful to P. Bussey, M. Derrick and T. Vaiciulis of the ZEUS
Collaboration for very helpful discussions. This work was funded in part 
by the US Department of Energy, Division of High Energy Physics, Contract No.
W-31-109-ENG-38.

\pagebreak


\pagebreak

\begin{table}
\caption{Total $\gamma + jet$ cross section in $pb$ with ZEUS cuts (see text).
\label{tab:exp}}
\vspace{0.4cm}
\begin{center}
\begin{tabular}{|r|r|r|r|r|l|}
\hline   
 &$standard$&$Q^2=(p_T^{\gamma})^2/4$&$Q^2=4(p_T^{\gamma})^2$&GRV$^{\gamma}$
                                                             &MRSR1\\
\hline
$res$&3.31  &2.60 &4.95 &6.72 &3.44   \\
\hline
$dir$&9.86 &11.45 &8.18 &9.86 &9.34 \\
\hline
$sum$ &$13.17$ &$14.05$ &$13.13$ &$16.58$ &$12.78$ \\
\hline
\end{tabular}
\end{center} 
\end{table}

\pagebreak

\noindent
\begin{center}
{\large FIGURE CAPTIONS}
\end{center}
\newcounter{num}
\begin{list}%
{[\arabic{num}]}{\usecounter{num}
    \setlength{\rightmargin}{\leftmargin}}

\item (a) Lowest order Feynman diagrams for the direct non-fragmentation
process $\gamma q \rightarrow \gamma q$. 
(b) Lowest order diagrams for the direct fragmentation
process $\gamma q \rightarrow g q$ and  $\gamma g \rightarrow q
\bar{q}$.

\item (a) Lowest order diagrams for the resolved non-fragmentation
process $q g \rightarrow \gamma q$ and  $q \bar{q} \rightarrow \gamma
g$. (b) Lowest order diagrams for two examples of the resolved
fragmentation processes.

\item (a) Rapidity distribution at fixed $p_T^\gamma$ for the inclusive 
non-isolated prompt photon cross section at HERA energies showing
resolved and direct contributions.
(b) Same as (a) but with the cut $0.16 \leq z \leq 0.8$ imposed on the
Weiszacker-Williams spectrum. 
(c) Same as (b) but showing the contributions from fragmentation
processes to both components as well as to the sum.
(d) Same as (b) and (c) but comparing the sum as calculated in LO and
NLO. 

\item Comparison of the non-isolated single photon and the isolated
photon plus jet cross sections at fixed $p_T^\gamma$ showing direct and
resolved contributions. The lower dot-dashed curve is the resolved
component after the cut $x_\gamma \geq 0.8$ is imposed.

\item The isolated photon plus jet cross section at $p_T^\gamma =5 $
GeV vs $\eta^\gamma$ with various cuts imposed by the ZEUS Collaboration
using (a) the GRV and GS96 photon structure functions and (b) the CTEQ4M
and MRSR1 proton structure functions.

\item $p_T^\gamma$ distribution of the resolved and direct contributions
to the photon plus jet cross section as well as their sum for photon rapidity 
in the range $-0.7\leq \eta^\gamma 0.8$.

\item $\eta^\gamma$ distribution of the photon plus jet cross section
for $0.5 \leq p_T^\gamma \leq 10$ GeV showing resolved (dot-dashed line)
and direct (dashed line) contributions and their sum (solid line). The
various dotted lines show the partial breakdown of the resolved
contribution. The dotted line with error bars is the contribution from
the $q^p g^\gamma$ initiated process using the GRV photon
parametrization.

\end{list}
\pagebreak
\epsffile{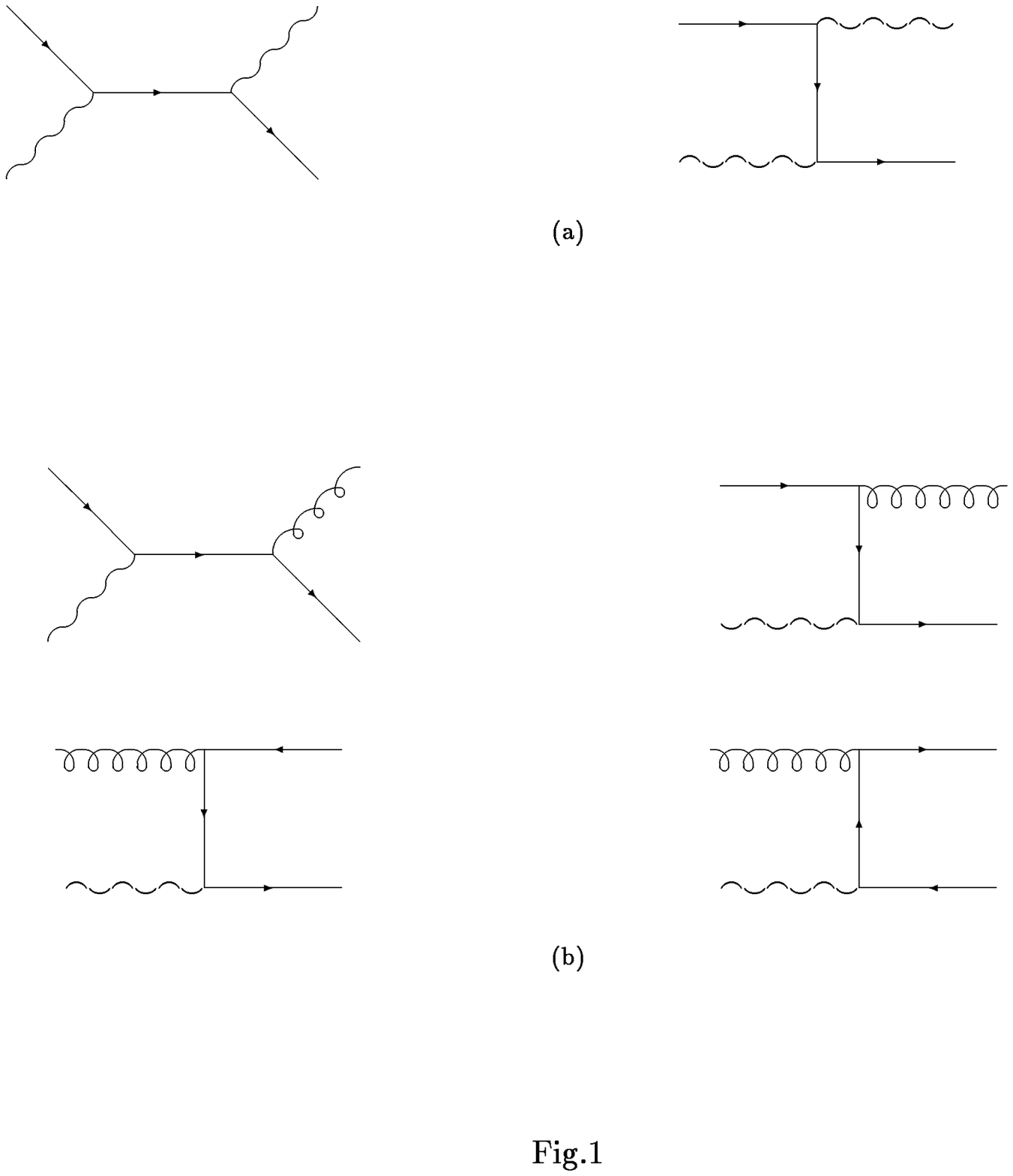}
\epsffile{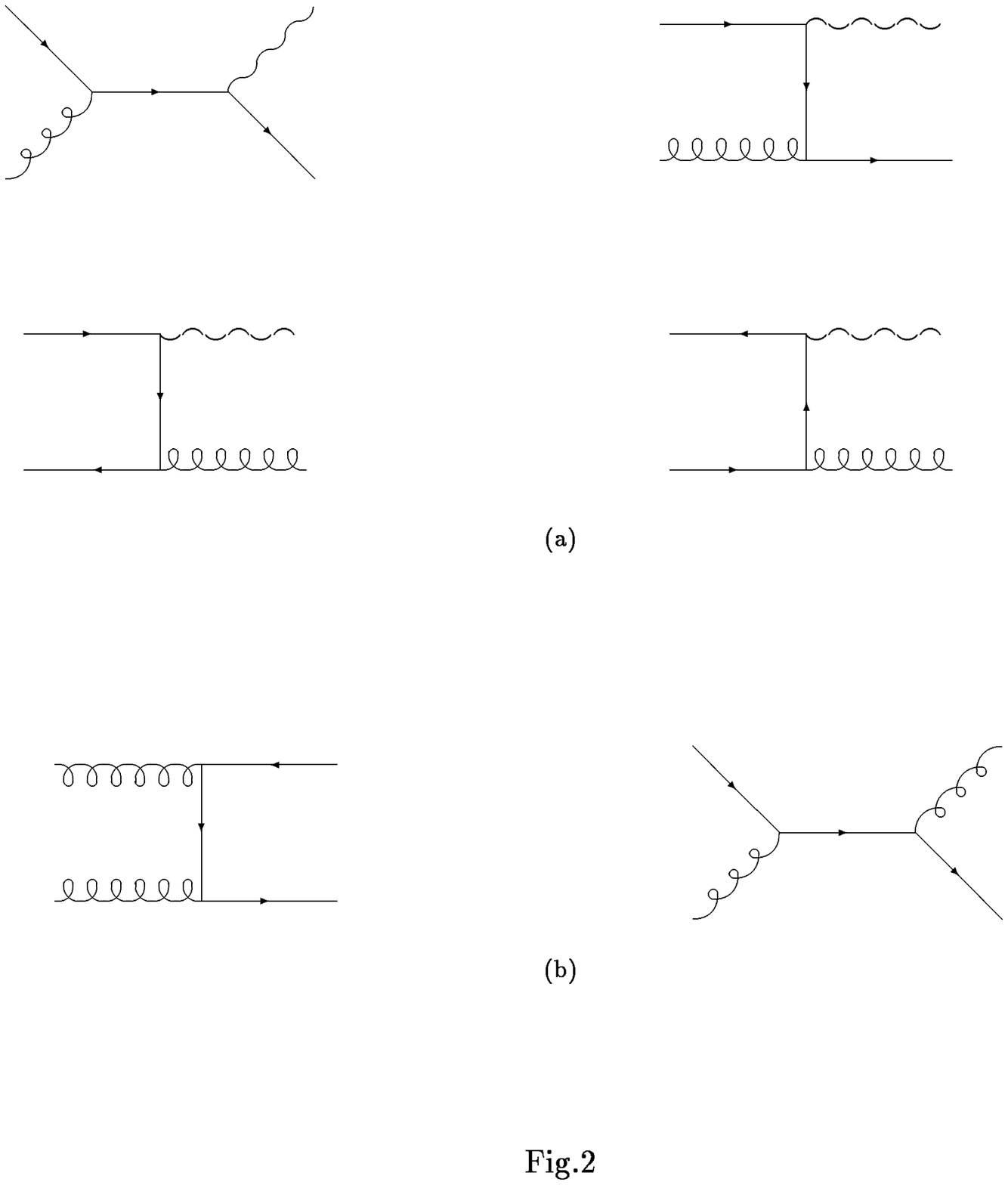}
\epsffile{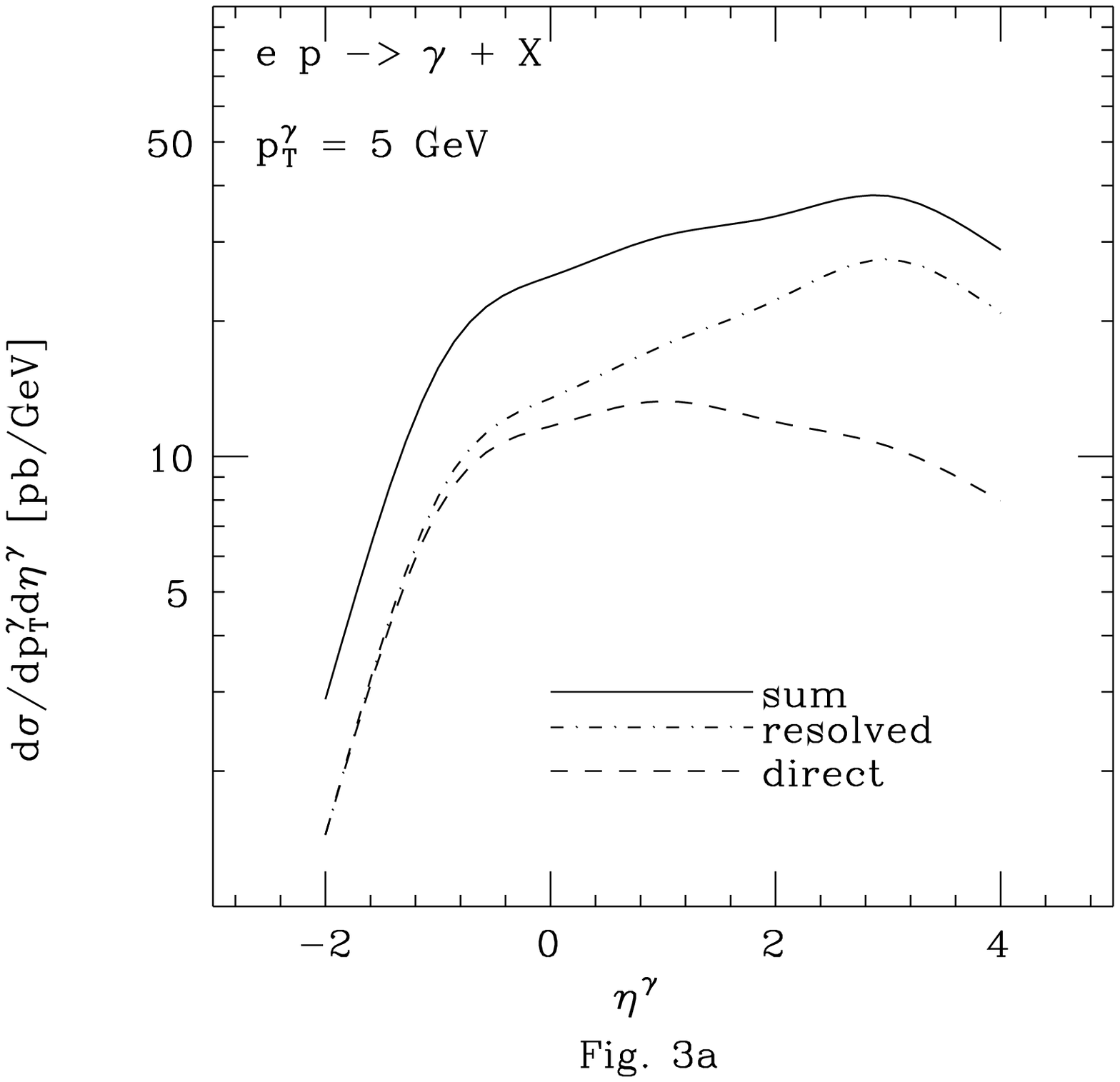}
\epsffile{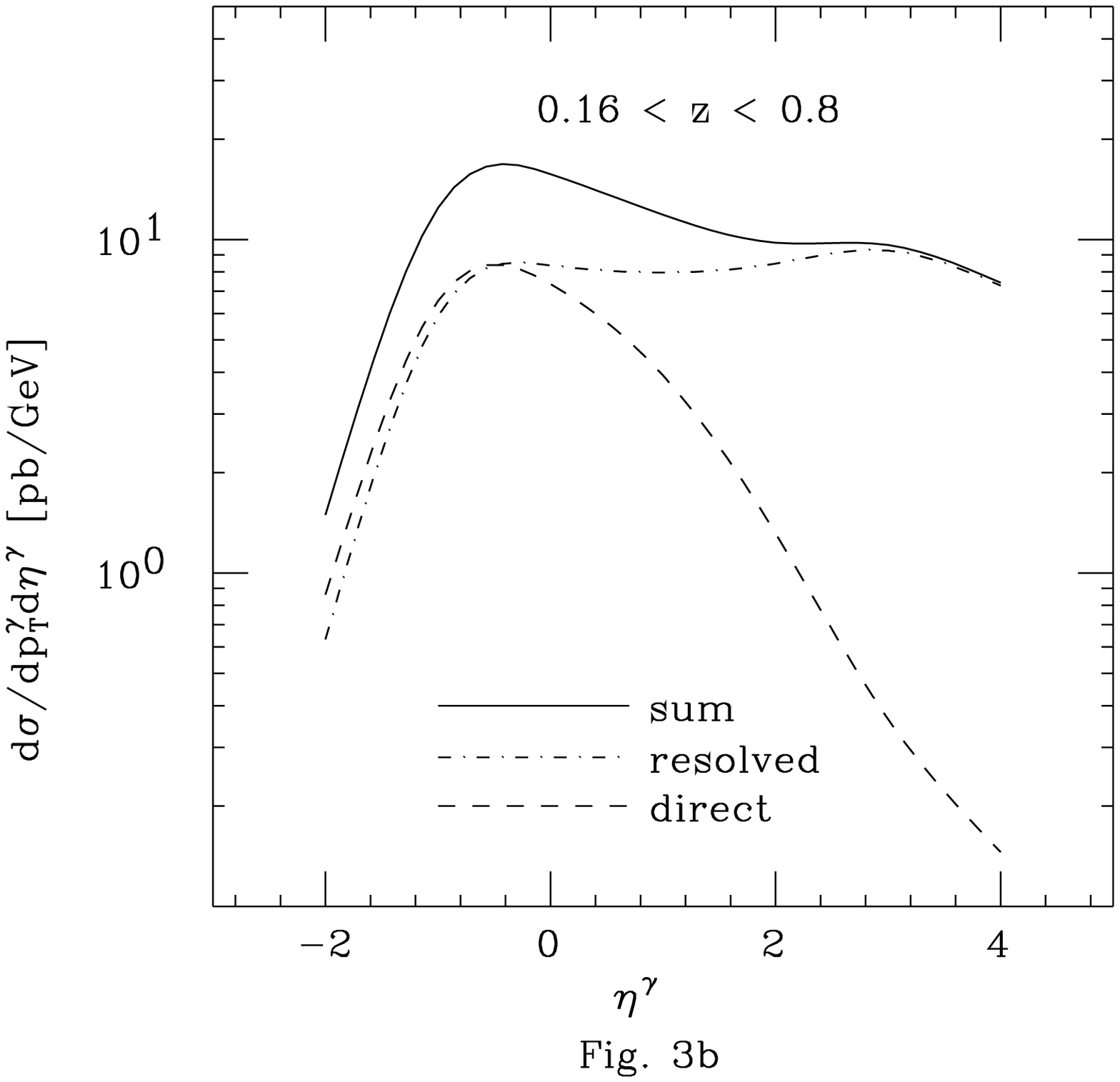}
\epsffile{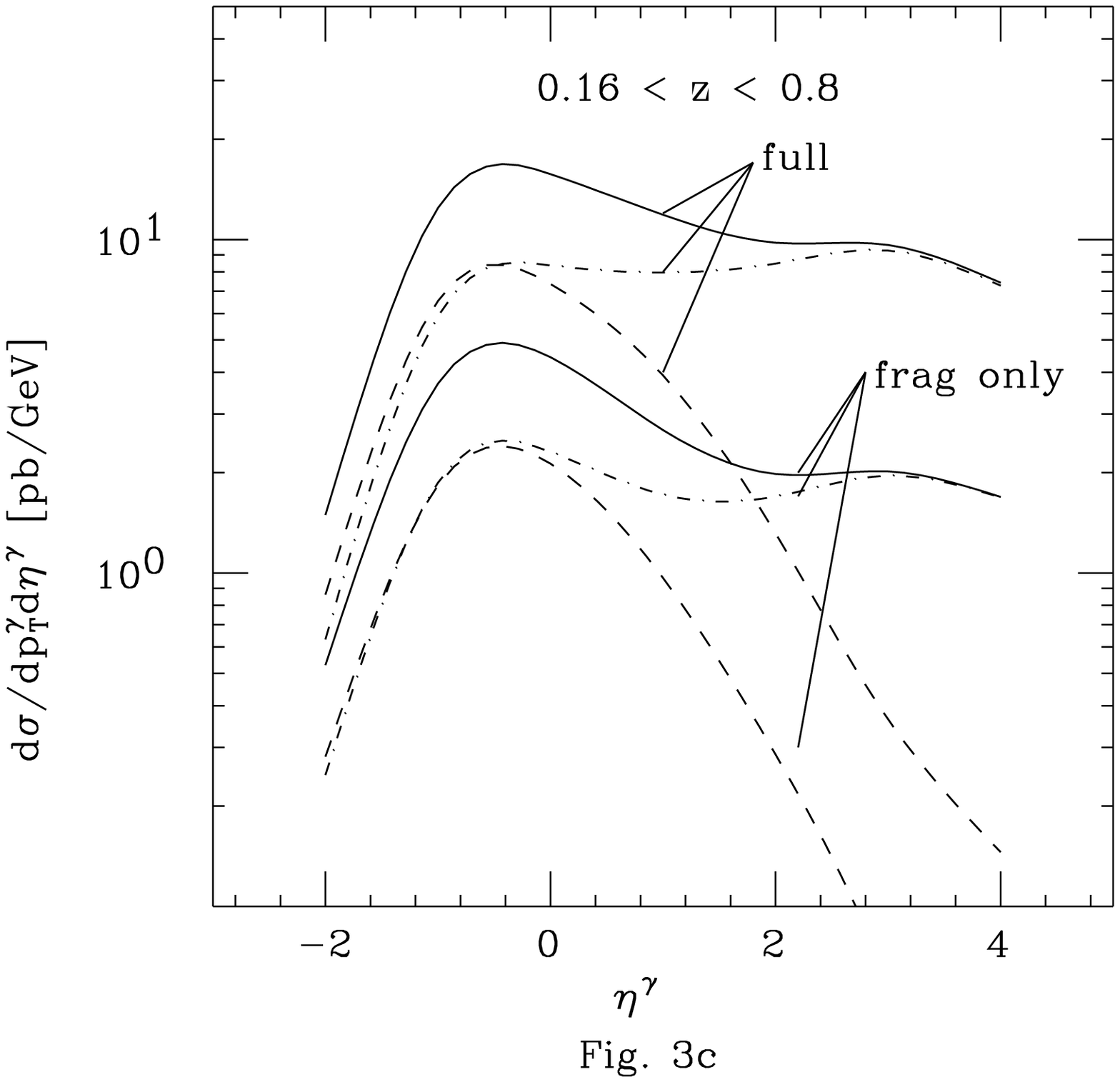}
\epsffile{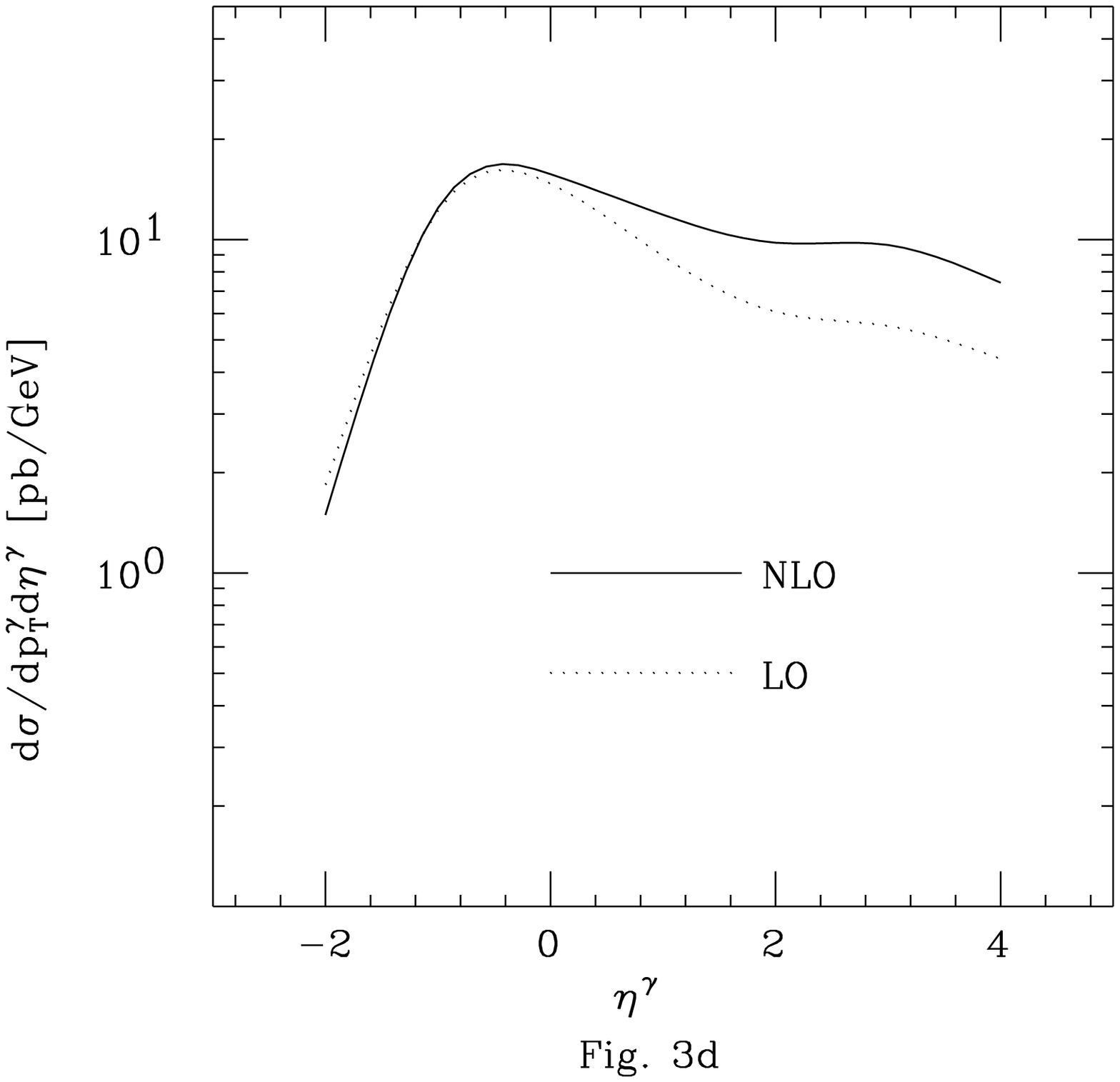}
\epsffile{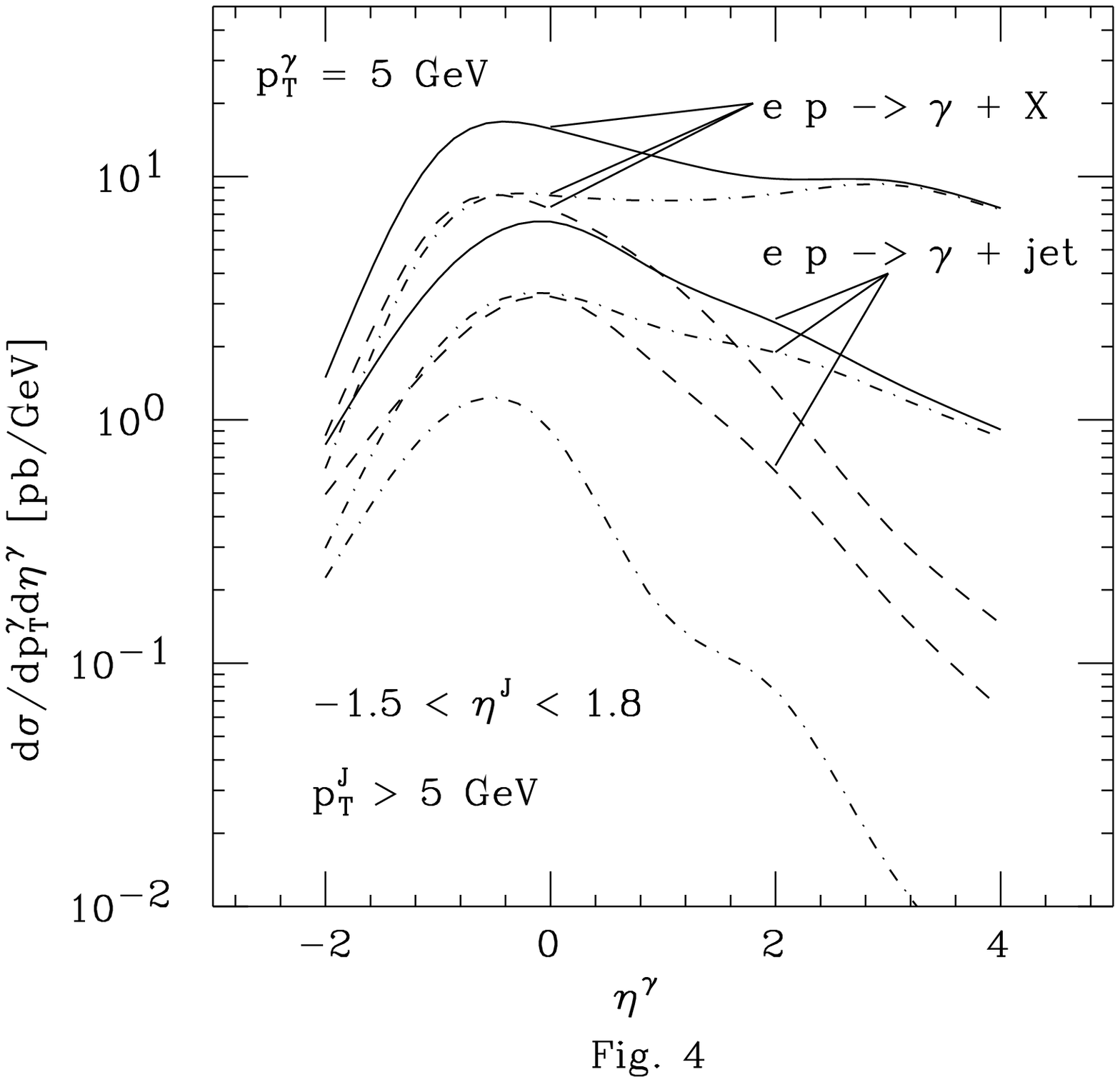}
\epsffile{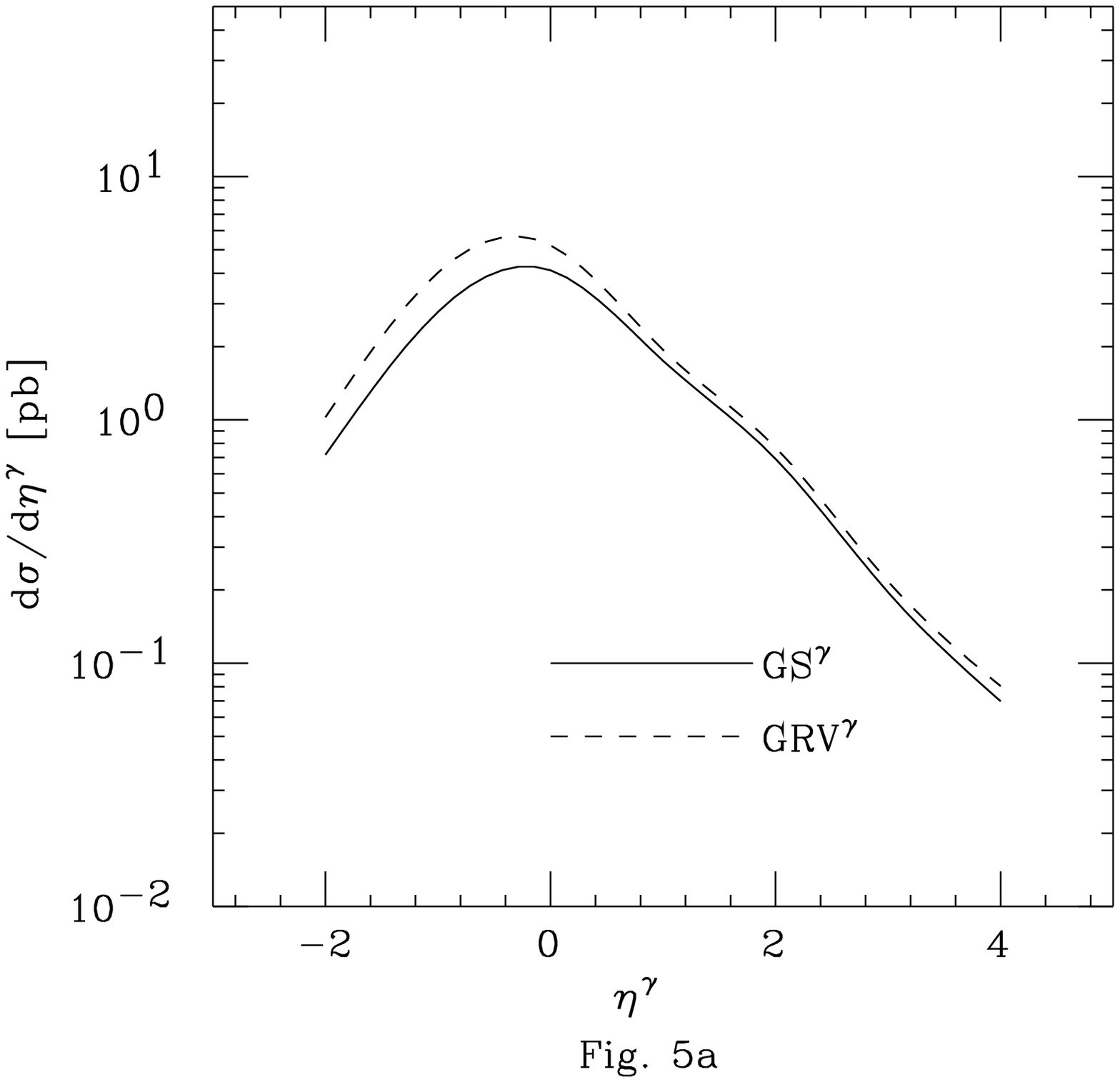}
\epsffile{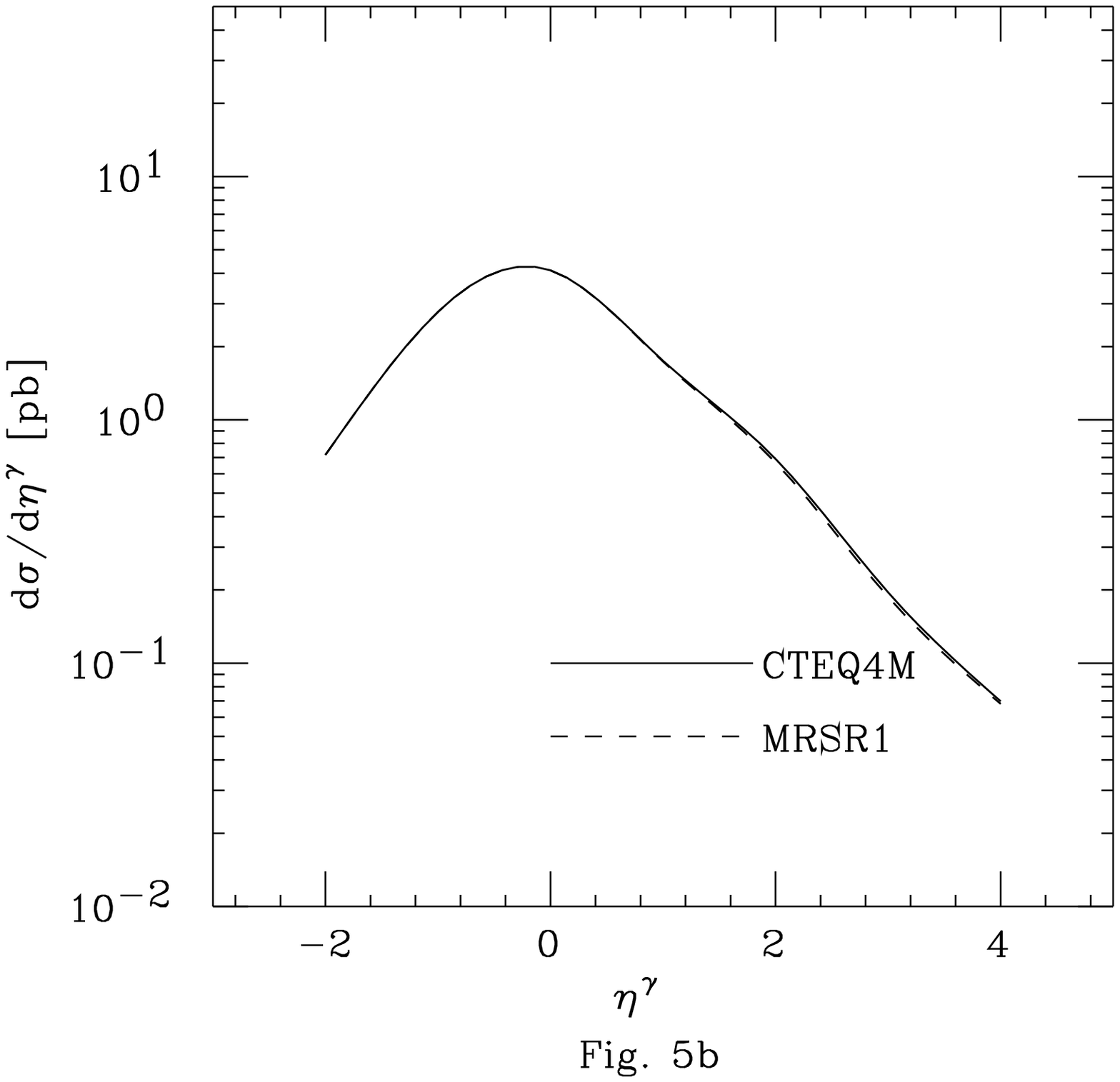}
\epsffile{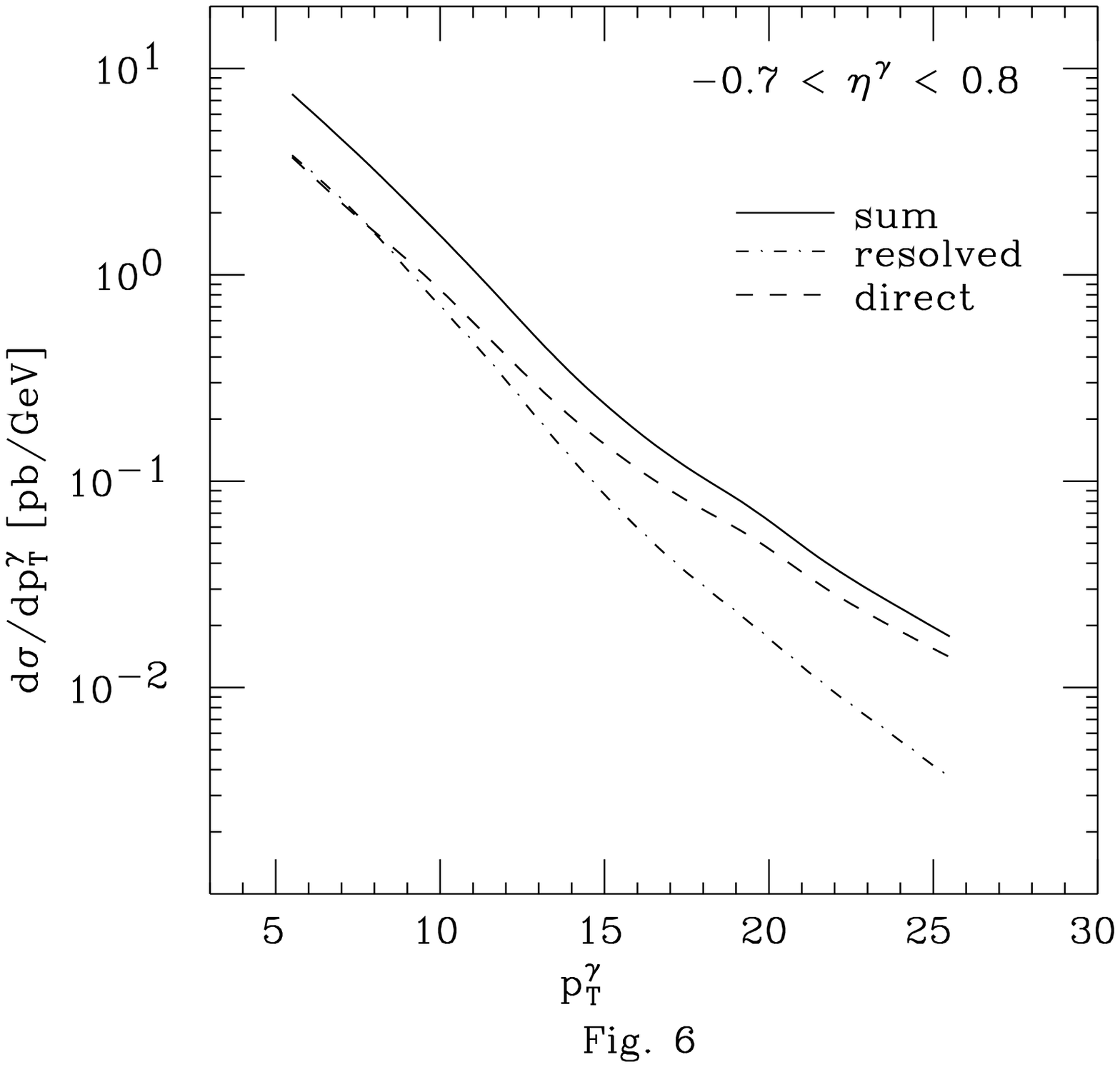}
\epsffile{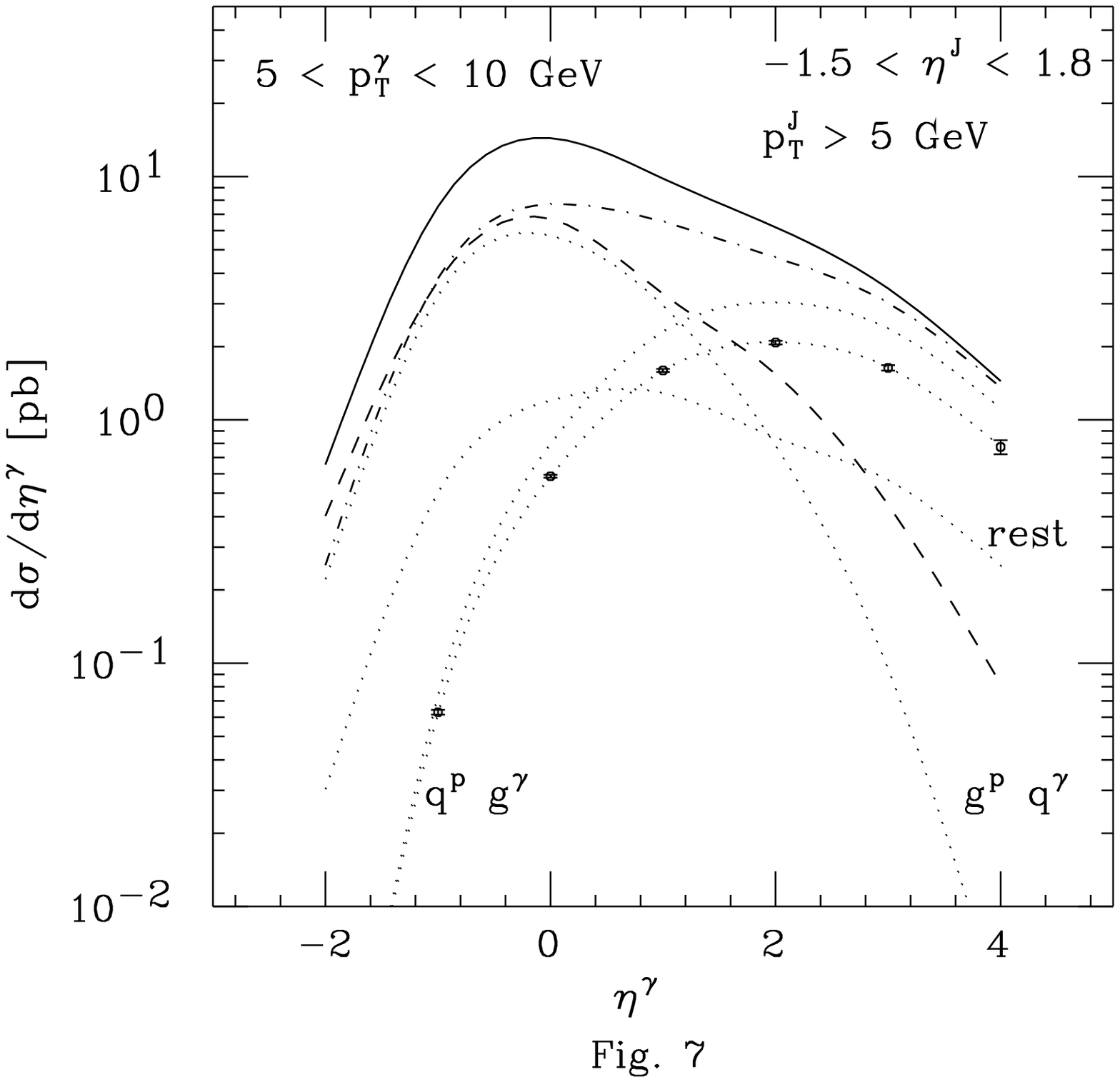}

\end{document}